\documentclass{nature}
\usepackage{graphicx}
\usepackage{float}
\usepackage{amsmath,amssymb,amsfonts}
\usepackage{bibentry}
\usepackage{lmodern}
\usepackage{hyperref}
\usepackage{color}
\usepackage{bm}
\usepackage{pgffor}
\usepackage{pdfpages}
\usepackage[normalem]{ulem}
\usepackage[font=small,labelfont=bf]{caption}
\makeatletter
\let\saved@includegraphics\includegraphics
\AtBeginDocument{\let\includegraphics\saved@includegraphics}
\renewenvironment*{figure}{\@float{figure}}{\end@float}
\makeatother

\spacing{1}

\hypersetup{
    colorlinks,%
    citecolor=black,%
    filecolor=black,%
    linkcolor=black,%
    urlcolor=black
}%

\newcommand{\Ree}{\mathop{\rm Re}\nolimits}

\newcommand{\Basel}{Department of Physics, University of Basel, Klingelbergstrasse 82, 4056 Basel, Switzerland.}
\newcommand{\FZUinst}{Institute of Physics (FZU), Czech Academy of Sciences, Na Slovance 2, 182 00 Prague 8, Czech Republic.}
\newcommand{\MFFinst}{Department of Condensed Matter Physics, Faculty of Mathematics and Physics, Charles University, Ke Karlovu 5, 121 16 Prague 2, Czech Republic.}
\newcommand{\Bern}{Department of Chemistry, Biochemistry and Pharmaceutical Sciences, W. In\"abnit Laboratory for molecular quantum materials, University of Bern, Freiestrasse 3, 3012 Bern, Switzerland.}

\author{Chao Li,$^{1,*}$ Vladislav Pokorn\'y,$^{2}$ Martin \v{Z}onda,$^{3,*}$ Jung-Ching Liu,$^{1}$ Ping Zhou,$^{4}$ Outhmane Chahib,$^{1}$ Thilo Glatzel,$^{1}$ Robert H\"aner,$^{4}$ Silvio Decurtins,$^{4}$ Shi-Xia Liu,$^{4,*}$ R\'emy Pawlak,$^{1,*}$ \& Ernst Meyer$^{1,*}$}

\title{Nanoscale Control of Quantum States in Radical Molecules on Superconducting Pb(111)}

\begin{document}
\maketitle
\begin{affiliations}
 \item\Basel
 \item\FZUinst
 \item\MFFinst
 \item\Bern

\noindent
{Correspondence to chao.li@unibas.ch, martin.zonda@matfyz.cuni.cz, shi-xia.liu@unibe.ch, remy.pawlak@unibas.ch, or ernst.meyer@unibas.ch}
\end{affiliations}

\section*{\Large {\bf Abstract}}
Magnetic impurities on superconductors present a viable platform for building advanced applications in quantum technologies. However, a controlled manipulation of their quantum states continues to pose a significant challenge, hindering the progress in the field. Here we show the manipulation of magnetic states in the radical molecule 4,5,9,10-tetrabromo-1,3,6,8-tetraazapyrene (TBTAP) on a Pb(111) superconducting surface using low-temperature scanning tunneling microscopy. Tunneling spectra reveal Yu-Shiba-Rusinov (YSR) states near the Fermi energy in isolated molecules. A quantum phase transition from singlet to doublet ground state is induced by changing the tip-molecule distance. Additionally, the presence of a second TBTAP molecule allows tuning of the YSR state position by altering the relative distance and can induce splitting of the YSR states for certain orientations. The construction of molecular chains up to pentamers shows periodic arrangements of charged and neutral molecules, with even-numbered chains forming a charged dimer structure at one end. Information can be encoded in these chains by switching the dimer position. These findings elucidate interactions between molecular assemblies and superconducting substrates, paving the way for advanced quantum-state engineering.

\section*{\Large \textbf{Main}}
Magnetic impurities adsorbed on a superconductor have shown great potential for exploring electron spin interactions with the surrounding electron environment\cite{heinrich2018single}. These systems are promising for advancing applications in spintronics\cite{linder2015superconducting}, quantum devices\cite{nadj2014observation}, and potentially enhancing the performance of superconducting diodes\cite{trahms2023diode}.
The impurities act destructively on Cooper pairs, breaking the time-reversal symmetry, leading to discrete states within the superconducting gap known as Yu-Shiba-Rusinov (YSR) states\cite{Balatsky-2006}. These states can be studied using scanning tunneling microscopy (STM) and spectroscopy (STS), where they appear in tunneling spectra as pairs of resonances symmetrically positioned around the Fermi energy\cite{heinrich2018single}. YSR states were observed in various systems, including individual magnetic adatoms\cite{ruby-2016,choi2017,cornils2017spin,kim2018toward,Odobesko2020,liebhaber2022quantum}, transition metal complexes\cite{kezilebieke2018coupled,malavolti2018tunable,shahed2022observation}, rare-earth molecular magnets\cite{Xia2022}, paramagnetic molecules\cite{homberg2020inducing,li2022surface} and metallic point contacts\cite{huang-2020,karan2024tracking}.

\noindent
Most studies focusing on spinful molecules use transition metal complexes in which the magnetic moment is mostly localized on the metal ion. Identifying a magnetic, metal-free molecule is a challenging task, but it could open up new possibilities for controlling spin behavior through novel methods. Significant efforts have been directed towards the exploration of organic magnets on metallic surfaces in recent years, stimulated by the growing interest in manipulating the characteristics of organic free radicals\cite{mas2012attaching,zhang2013temperature,sun2020inducing,he2022observation}. Recently, a $D_{2h}$-symmetric molecule 4,5,9,10-tetrabromo-1,3,6,8-tetraazapyrene (TBTAP) (Fig.~\ref{Fig1}a), has been reported to have a spin-1/2 magnetic ground state when adsorbed on Ag(111) or Pb(111) surfaces\cite{li2023strong,liu2023gate}. This state is formed by occupation of the TBTAP LUMO orbital with an electron donated by the substrate. The TBTAP molecule holds promise as a candidate for a building block of nanoscopic assemblies for molecular technologies.

\noindent
Here, we present a low-temperature STM and STS study of both individual TBTAP$^{\bullet-}$ radical molecules and  molecular chains on Pb(111) superconducting surface. We observed an impurity quantum phase transition (QPT) between a singlet and a doublet ground state governed by the distance between the superconducting STM tip and an isolated molecule. Furthermore, we achieved control over the magnetic state of molecular dimer by manipulating the relative distance and orientation. The experimental findings are in agreement with theoretical predictions based on density functional theory (DFT) calculations and on the solution of the superconducting impurity Anderson model, which provides a reliable explanation for all observed phenomena. Chains of TBTAP molecules from three to five units long were created and studied for their ability to form systems consisting of a combination of charged and neutral molecules. We proved that information can be encoded in tetramer chains by switching their charge state between two equivalent configurations.

\section*{\Large \textsf{Morphology and tunneling spectra characteristics of individual TBTAP molecules on Pb(111).}}
\begin{figure}[!ht]
    \begin{center} \includegraphics[width=0.5\columnwidth]{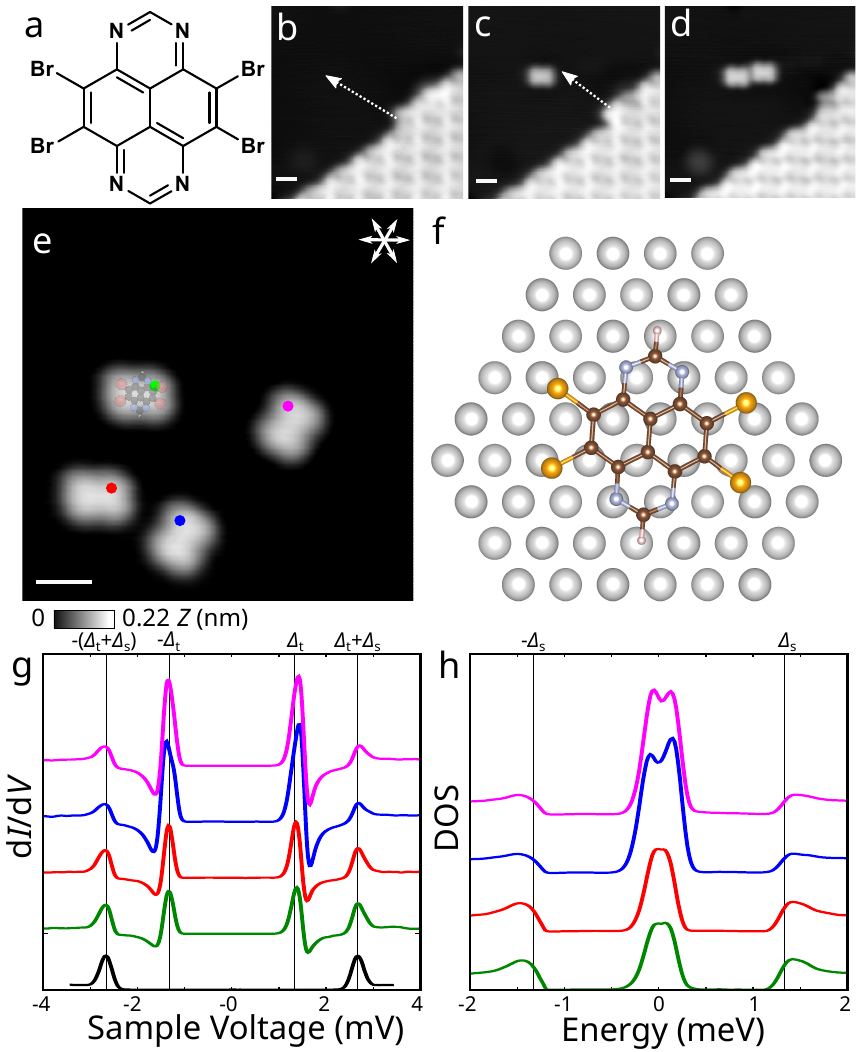} \end{center}
    \caption{\textbf{Manipulation and YSR states of isolated molecules.}
    ({\bf a}) Chemical structure of the 4,5,9,10-tetrabromo-1,3,6,8-tetraazapyrene (TBTAP) molecule. 
    ({\bf b} - {\bf d}) Creation of isolated TBTAP molecules by relocation from a molecular island through lateral manipulations ($I$= 100 pA, $V_{\rm s}$= 500, 300 and 100 mV for {\bf b}, {\bf c} and {\bf d}, respectively).
    ({\bf e}) Four isolated TBTAP molecules exhibit preferred orientations aligned with densely packed directions of the surface with a small deviation of $4^\circ\pm1^\circ$ ($I$= 100 pA, $V_{\text{s}}$= 100 mV).
    ({\bf f}) Equilibrium position of a TBTAP molecule on Pb(111) surface calculated using DFT. The molecule is oriented along the densely packed directions of the surface with a deviation of $4.2^\circ$ between the $\langle 110\rangle$ surface direction and one of the $c_2$ axes of the molecule, in very good agreement with the experiment.
    ({\bf g}) $dI/dV$ spectra of isolated molecules measured at the positions marked by dots in panel {\bf e}. In addition to the superconductor gap edge at $\Delta_{\text{t}}+\Delta_{\text{s}}=2.6$\!~meV, 
    clear additional resonances appear inside the gap at $eV$ close to $\Delta_{\text{t}}$, indicating the presence of YSR states in isolated molecules. We added a spectrum for the pure Pb surface (black line) for comparison. The spectra have been vertically shifted for clarity. ($V_{\text{s}}$ = 5 mV; $I_{\text{t}}$ = 0.4 nA, $A_{\text{mod}}$ = 0.05 mV, $f$ = 613 Hz).
    ({\bf h}) Surface DOS of isolated molecules obtained by the deconvolution procedure from the tunneling spectra. YSR states lie very close to the Fermi energy, suggesting that the system is close to a QPT. The tip parameters for the deconvolution read $\Delta_{\text{t}}=1.31$\!~meV and $\gamma_{\text{t}}=0.04$\!~meV. The scale bars in all figures represent 1 nm. 
    }
    \label{Fig1}
\end{figure}

\noindent
TBTAP molecules have recently been reported to exhibit a radical TBTAP$^{\bullet -}$ state within self-assembled monolayers on Pb(111)\cite{liu2023gate}. Such radical has a ground state $S=1/2$ that gives rise to YSR states within the superconducting gap. To study the intrinsic behavior of individual TBTAP molecules, they were removed by the STM tip from a molecular island (Fig.~\ref{Fig1}b-d). Each molecule is adsorbed along the close-packed surface direction $\langle 110 \rangle$, with a small deviation of $4^\circ\pm 1^\circ$. This behavior is in good agreement with DFT result (Fig.~\ref{Fig1}f), which predicts the molecule to lie $\approx 330$\!~pm above the surface at the angle of $4.2^\circ$. The magnetic state was verified by measurements using a gradually increasing out-of-plane magnetic field up to $1$\!~T at $T_{\text{exp}}=2.2$\!~K. The magnetic field used is strong enough to suppress the superconductivity in the tip and the substrate. The STS results show a well-developed Kondo peak. Its Frota fit\cite{frota92} gives an estimated Kondo temperature $T_\text{K}^\text{FF}=7.5$\!~K (see Supplementary Note~4).

\noindent
Differential conductance ($dI/dV$) spectra were recorded at $T_{\text{exp}}=1$\!~K on four molecules at positions marked by dots in Fig.~\ref{Fig1}e. Due to the use of a superconducting Pb tip, the tunneling spectra are convolutions of the density of states (DOS) of the tip (t) and of the substrate (s) and all energies are shifted by the tip gap $\Delta_{\text{t}}$. The spectra are plotted in Fig.~\ref{Fig1}g, together with a spectrum of a bare Pb surface. In addition to the coherence peaks at $\Delta_{\text{t}}+\Delta_{\text{s}}=\pm2.6$\!~meV, a single pair of slightly asymmetric peaks emerge at $V=\pm1.4$\!~mV. The asymmetry of the peaks is due to the Coulomb potential which breaks the particle-hole symmetry\cite{Balatsky-2006} and is a hallmark of the single-electron nature of the involved tunneling process\cite{ruby2015}. The character of the YSR states is better illustrated on the surface DOS obtained from the tunneling spectra by deconvolution (see Supplementary Note~2) plotted in Fig.~\ref{Fig1}h, which shows the YSR states that lie close to the Fermi energy. This behavior indicates that the system is close to an impurity QPT between the singlet and doublet ground states\cite{defranceschi2010hybrid}.

\noindent
Fig.~\ref{Fig2}a displays a local density map of a molecule at the YSR position of $1.4$\!~mV (see Supplementary Note~1). This map unveils the spatial distribution of the YSR state, which has the same symmetry as the spin density of a TBTAP$^{\bullet -}$ calculated using DFT (Fig.~\ref{Fig2}b). The YSR states are delocalized over the backbone of aromatic rings. This is in contrast to experiments performed on organometallic complexes, where the YSR states are localized on the metallic ion. However, no coherent long-range extent of the YSR state away from the molecule was observed. We attribute this to the three-dimensional nature of the Pb(111) substrate, in which the spatial extent of the YSR states is strongly suppressed compared to experiments conducted on quasi-two-dimensional substrates such as NbSe$_2$\cite{menard2015}. 

\section*{\Large \textsf{Quantum phase transition in an individual TBTAP molecule.}}
\begin{figure}[!ht]
    \begin{center} \includegraphics[width=0.9\columnwidth]{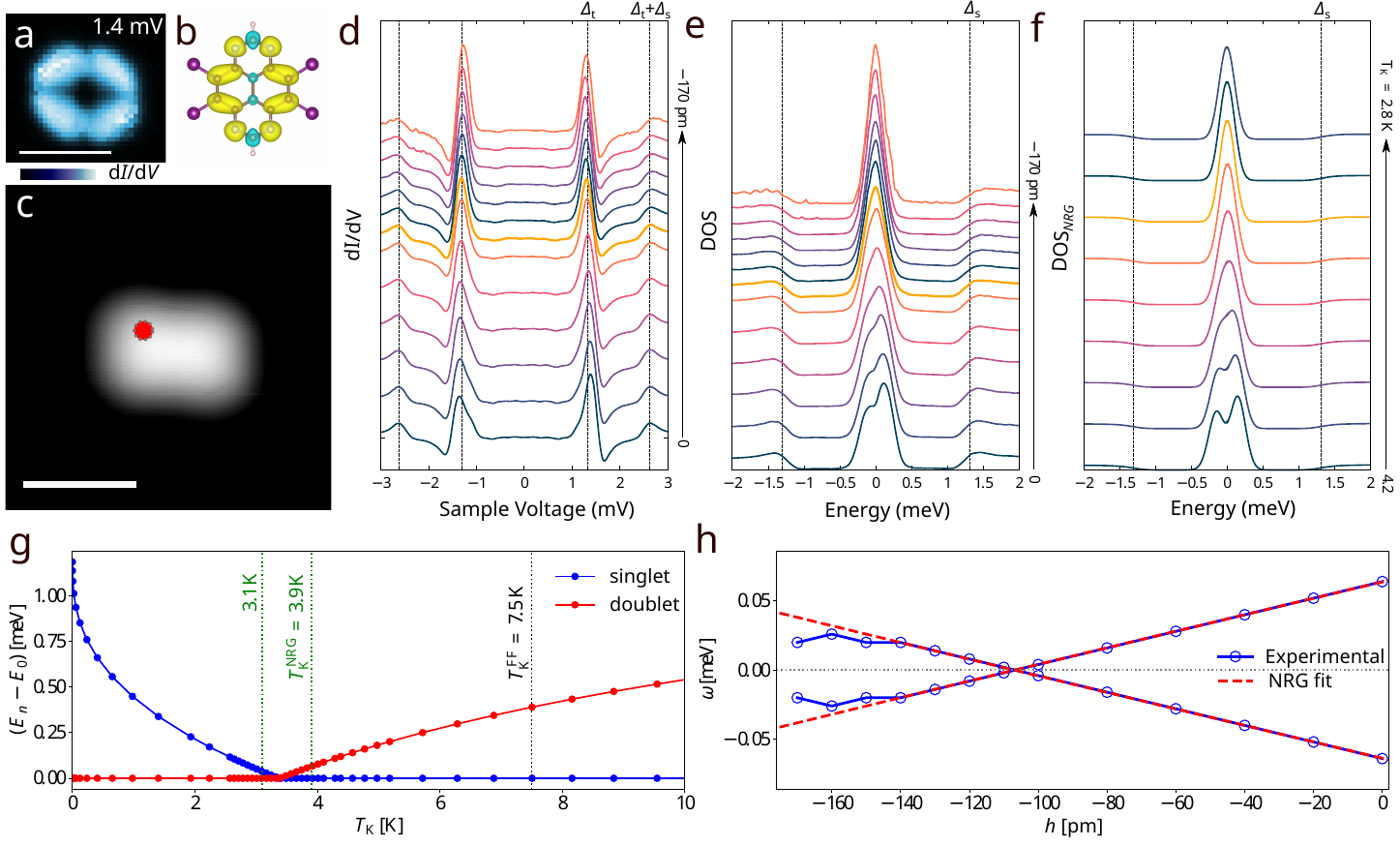} \end{center}
    \caption{\textbf{Quantum phase transition in individual molecules.}
    ({\bf a}) Grid map of $dI/dV$ at $V=1.4$\!~mV, close to the YSR resonance.
    ({\bf b}) Spin density map of the charged TBTAP molecule on the Pb substrate calculated using DFT.
    ({\bf c}) A high resolution STM image ($V_{\text{s}}=0.1$\!~V and $I_{\text{t}}=0.1$\!~nA) of an individual TBTAP molecule.    
    ({\bf d}) Differential conductance spectra at different tip heights $h$ measured using the superconducting Pb tip. The initial tip position was set at $I$= 100 pA, $V_{\text{s}}$= 5 mV above the molecule, followed by a gradual approach towards the molecule. The Fermi energy value was corrected in each data set by a small shift determined from the positions of the coherence peaks.
    ($V_\text{s}$ = 6 mV, $A_\text{mod}$ = 0.02 mV, $f$ = 613 Hz).
    ({\bf e}) Surface DOS obtained using the deconvolution procedure from $dI/dV$ data in panel {\bf d}.
    The thicker yellow lines in panels {\bf d} and {\bf e} represent data for $h=-110$\!~pm, closest to the QPT. The tip parameters for the deconvolution read $\Delta_{\text{t}}=1.31$\!~meV and $\gamma_{\text{t}}=0.04$\!~meV.
    ({\bf f}) Surface DOS calculated using NRG for different values of the Kondo temperature $T_K$ for $U=200$\!~meV and slightly away from half-filling ($\varepsilon=-20$\!~meV) to simulate the asymmetry in the differential conductance spectra.
    ({\bf g}) Subgap states, respectively the differences between the first excited state and the ground state energies, of the SC-SIAM as functions of $T_\text{K}$ calculated using NRG. The blue (red) points represent the singlet (doublet) state. The black dashed line marks the $T_\text{K}$ obtained from the fit of the normal-state Kondo peak with the Frota function. The green dashed lines mark the range of $T_\text{K}$ corresponding to the change of the vertical distance of the scanning tip from the molecule, fitted from the position of the YSR peaks in $dI/dV$ via SC-SIAM solved by NRG.
    ({\bf h}) Positions of the subgap state maxima extracted from the experimental $dI/dV$ data and shifted by $\Delta_{\text{t}}$ (blue symbols). The red dashed line represents the fit by NRG as described in the text. The energies were averaged over the positive and negative values to increase the quality for small distances.
    }
    \label{Fig2}
\end{figure}

The YSR states are close to the Fermi energy which implies that the system is naturally set in the vicinity of a QPT. This transition can be achieved by different means, including using different adsorption sites\cite{franke-2011}
or applying mechanical forces with an STM tip\cite{Farinacci-2018}. To demonstrate the  transition we performed a series of STS measurements on the molecule while varying the tip height at the position marked by a red dot in Fig.~\ref{Fig2}c. The controllable motion of molecules during lateral manipulation suggests the existence of an attractive force between the molecule and the tip. This force lifts the molecule from the substrate and has the potential to fine-tune the coupling strength between the molecule and its environment. We approached the scanning tip within $170$\!~pm of its initial position ($I=100$\!~pA, $V=10$\!~mV). The resulting spectra are plotted in Fig.~\ref{Fig2}d. The YSR position decreases until $h\approx-110$\!~pm and then increases again, suggesting the crossing of YSR states. To better visualize this crossing, we performed a deconvolution to obtain the surface DOS (Fig.~\ref{Fig2}e). The position of the YSR state changes from $60$\!~$\mu$eV to -30\!~$\mu$eV, indicating a QPT from a singlet ground state at large distances to a doublet ground state at small distances.

\noindent
This behavior can be described using the superconducting single impurity Anderson model (SC-SIAM)\cite{meden2019anderson}, which is known to provide reliable insights into the physics of superconducting impurity systems\cite{Luitz-2012}. We describe the system as a single quantum level tunnel-coupled to two superconducting leads, the surface and the tip. More details are provided in the Supplementary Note~3. From now on, we assume that $\Delta_{\text{t}}=\Delta_{\text{s}}=\Delta$. The decrease in tip height increases the coupling to the tip $\Gamma_{\text{t}}$, but decreases the coupling to the substrate $\Gamma_{\text{s}}$ by lifting the molecule. As $\Gamma_{\text{s}}\gg\Gamma_{\text{t}}$ this leads to a decrease in total coupling $\Gamma=\Gamma_{\text{s}}+\Gamma_{\text{t}}$. Note that at small distances the force between the tip and the molecule eventually changes to repulsive, again leading to an increase in $\Gamma_{\text{s}}$\cite{Farinacci-2018}.

\noindent
The governing scale of the singlet-doublet QPT in SC-SIAM for the parameter regime of the experiment is the Kondo temperature $T_\text{K}$\cite{Kadlecova-2019,meden2019anderson}. The ground state changes from a singlet to a doublet at the critical value $T^c_\text{K}$ that can be approximated by\cite{Kadlecova-2019} $k_\text{B}T^c_\text{K} \approx \Delta/(e^{5/3}-1)\doteq0.23\Delta$. For $\Delta=1.31$\!~meV this gives $T^c_\text{K}= 3.5$\!~K. The small YSR energies suggest that the system is close to the QPT already at the initial position of the tip. Therefore, we expect that $T_\text{K}$ is close to $T^c_\text{K}$ in the whole measured range of the tip height.
To test this conjecture, we have analyzed the experimental data using numerical renormalization group\cite{zitkoNRGljubljana} (NRG) solution of SC-SCIAM (see Methods), which provides us with an alternative way to extract the $T_\text{K}$ of the setup. Fig.~\ref{Fig2}g shows the energy of the subgap states with respect to the ground state energy $E_0$ as functions of $T_\text{K}$ for a half-filled impurity. The differences $\omega^\pm=\pm|E_1-E_0|$ are the YSR energies, which appear as subgap peaks in the surface DOS. YSR states cross at the Fermi energy at $T_\text{K}^c\simeq 3.5$~K as expected from the above estimate. 

\noindent
The YSR energy extracted from the tunneling spectra as the maximum of the in-gap peak as a function of the tip distance $h$ is shown in Fig.~\ref{Fig2}h. Note that $h=0$ is just a reference value that refers to $650\pm 50$\!~pm above the molecule. The YSR energy depends almost linearly on $h$ up to very small distances, where the YSR states turn again toward the Fermi energy. This behavior is a result of the force between the tip and the molecule changing from attractive to repulsive, pushing the system back toward the QPT. We can use values in the linear regime around the QPT to obtain a better estimate of $T_\text{K}$ by an NRG fit. However, we first need a mapping between $T_\text{K}$ and $h$.
To that end, we utilized the Wilson's formula for the Kondo temperature, Eq.~\eqref{eq:TK}, together with a linearization procedure explained in the Supplementary Note~5. In the vicinity of QPT this leads to a linear relation $h=(\Gamma-\Gamma_0)/\tilde{\alpha}$, where $\Gamma_0$ and $\tilde{\alpha}$ are parameters that can be fitted from the experimental data. The NRG results, marked by the dashed red lines in Fig.~\ref{Fig2}h, overlap with the experiment and can be used to extract $\Gamma_\text{exp}$ and consequently $T_\text{K}$. Assuming $U=200$~meV\cite{li2023strong,liu2023gate}, we find that $\Gamma_\text{exp}$ changes between $18.2$\!~meV and $19.8$\!~meV in the linear regime that gives $T^\text{NRG}_\text{K}$ between $3.1$\!~K and $3.9$\!~K.

\noindent
In Fig.~\ref{Fig2}f we show the surface DOS calculated using NRG for values of $T_\text{K}$ from the extracted interval. The evolution of DOS is in agreement with the experimental results. The difference in weights of the bands above the gap between theory and experiment can be attributed to an experimental background not included in the SC-SIAM analysis. 

\noindent
Estimation of $T_\text{K}$ for the single-molecule system revealed an inconsistency between the Frota fit and the NRG result. There are multiple mechanisms that can lead to this disagreement, e.g., NRG relies on estimates of model parameters such as Coulomb repulsion. However, due to Kondo universality, the dependence of YSR state on $T_\text{K}$ from Fig.~\ref{Fig2}g and consequently also the NRG fit are very robust when varying the model parameters within experimentally meaningful limits.
On the other hand, it has already been shown\cite{zonda2021resolving} that a simple Frota fit of the Kondo peak significantly overestimates the Kondo temperature when $T_\text{K}$ is comparable to the experimental temperature, which is our case as $T_{\text{exp}}=2.2$\!~K for the normal state measurement. Therefore, we argue that NRG fit gives a better estimate of the real Kondo temperature than Frota and that the system is indeed naturally very close to a QPT.

\noindent
\section*{\Large \textsf{Manipulating YSR states in molecular dimers.}}

\begin{figure}[!ht]
    \begin{center} \includegraphics[width=0.9\columnwidth]{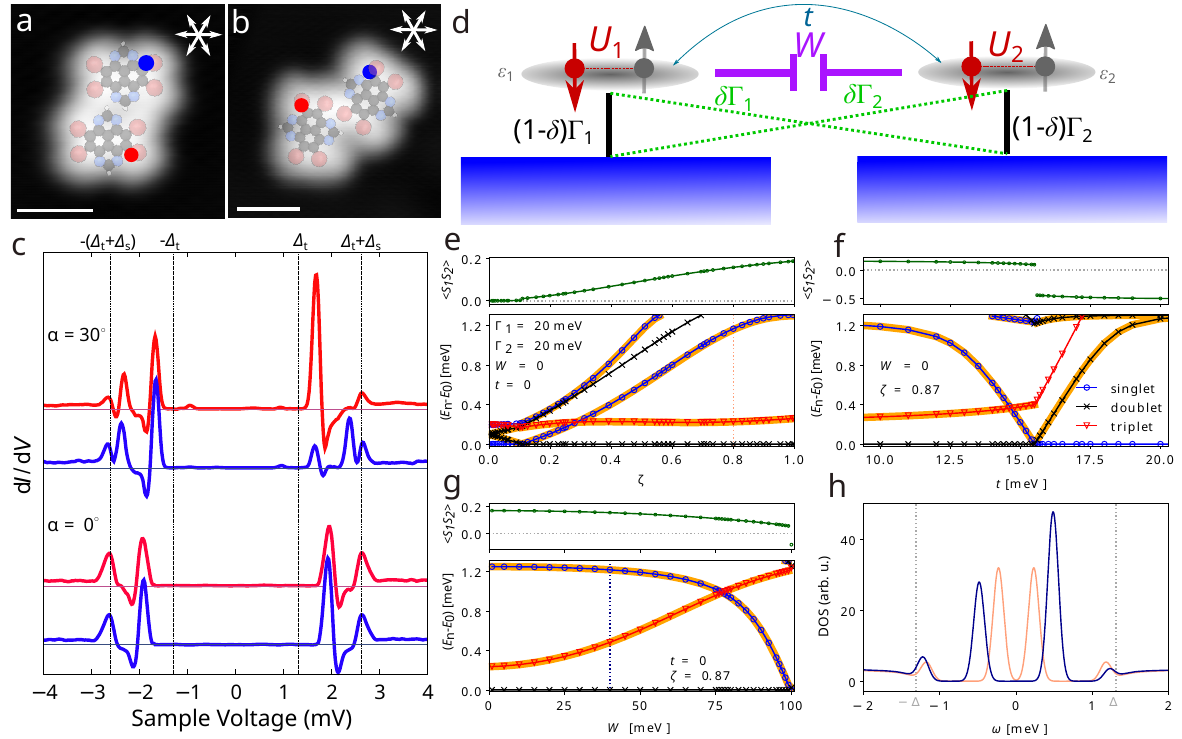} \end{center}
    \caption{\textbf{Molecular dimers.}
    ({\bf a} - {\bf b}) Two configurations of molecular dimers constructed by lateral manipulations 
    ($V_\text{s}=0.1$\!~V and $I_\text{t}=60$\!~pA for {\bf a}; 
    $V_\text{s}=50$\!~mV and $I_\text{t}=0.1$\!~nA for {\bf b}).
    ({\bf c}) Tunneling spectra of molecules from panels {\bf a} ($\alpha=0^\circ$) and {\bf b} ($\alpha=30^\circ$) measured at positions marked by the red and blue dots. The spectra have been vertically shifted for clarity.
    ({\bf d}) Illustration of the possible couplings between two correlated impurities described by SC-TIAM. For $\delta=0$ the two impurities act as decoupled and we obtain a model for large inter-molecular distances; for $\delta=1/2$ we end up with a model for a local two-level impurity. Furthermore, $t$ denotes the amplitude of a direct hopping between the impurity orbitals and $W$ the strength of inter-impurity capacitive coupling.
    ({\bf e} - {\bf g}) Bottom panels: YSR states of a dimer described by SC-TIAM calculated using NRG as functions of the inter-molecular coupling parameter $\zeta$ (panel {\bf e}), direct hopping $t$ (panel {\bf f}) and capacitive coupling $W$ (panel {\bf g}), respectively. The blue circles, black crosses, and red triangles denote the singlet, doublet, and triplet states, respectively. Orange lines denote the actual YSR states which appear in the STS data, that is, transitions that are not forbidden by the $\Delta s_z=\pm1/2$ parity selection rule. Top panels: The respective inter-impurity spin-spin correlation function $\langle \bm{S}_1\bm{S}_2\rangle$ in units of $\hbar^2$. Positive values signal a ferromagnetic effective RKKY exchange, and negative values mark an antiferromagnetic order, where electrons form an entangled singlet state. 
    ({\bf h}) Surface DOS calculated using NRG for the same parameters as in panel {\bf e} for $\zeta=0.8$ and $W=0$ (salmon) and panel {\bf g} $\zeta=0.87$ and $W=40$\!~meV (dark blue). This value of the inter-impurity coupling shifts the YSR energies close to the experimental values in the upper part of panel {\bf c} ($\alpha=30^\circ$).
    } 
    \label{Fig3}
\end{figure}

\noindent
Another way to manipulate YSR states in molecular systems is by engineering setups consisting of two impurities. Two different types of molecular dimers were constructed using lateral manipulation technique. A density map of a dimer consisting of molecules with the same orientation is shown in Fig.~\ref{Fig3}a. This configuration is stabilized by non-covalent interactions such as N$\cdots$H hydrogen bonds. Tunneling spectra, recorded at positions marked by the dots in Fig.~\ref{Fig3}a, are plotted at the bottom of Fig.~\ref{Fig3}c ($\alpha=0^\circ$) by respective colors. Both molecules exhibit almost identical spectra that show a single pair of YSR states at $V=\pm1.9$\!~mV. 
The evolution of the position of the peaks between the limit of two isolated molecules and the dimer is briefly discussed in the Supplementary Note~8. 

\noindent
Another configuration of molecules is shown in Fig.~\ref{Fig3}b. In this configuration, stabilized by Br$\cdots$Br and Br$\cdots$N halogen bonds, the $c_2$ axes of the two molecules form a $30^\circ$ angle. It should be noted that the orientation of one molecule in this dimer is rotated by $30^\circ$ compared to the isolated molecule, due to the change in adsorption orientation caused by the adjacent molecule. The corresponding tunneling spectra, plotted in the upper part of Fig.~\ref{Fig3}c ($\alpha=30^\circ$), exhibit two pairs of YSR states at $V_1=\pm1.7$\!~mV and $V_2=\pm2.5$\!~mV. The deconvolved surface DOS for both configurations can be found in the Supplementary Note~8. 

\noindent
We analyzed the behavior of the dimer system using the superconducting two-impurity Anderson model (SC-TIAM). Its schematic representation is shown in Fig.~\ref{Fig3}d and a detailed description can be found in Supplementary Note~6. In general, the effect of the distance between impurities is encoded in SC-TIAM via complex hybridization terms\cite{eickhoff-2018}. However, for a constant substrate DOS this effect can be parameterized by tunneling cross-terms (green lines in Fig.~\ref{Fig3}d). Here, $\delta\in\langle0,1/2\rangle$, respectively $\zeta=2\sqrt{(1-\delta)\delta}$ (see Supplementary Note~6), is a parameter that indirectly encodes the inter-molecular distance. For $\delta=0$ we have a system of two distant impurities effectively coupled to their own substrates. The other limit $\delta=1/2$ describes a single two-level impurity coupled to a single substrate\cite{yao2014phase,zalom2024double}. This way, $\delta$ can be used to tune the inter-molecular correlations through the substrate that can lead to both effective coupling and effective exchange interaction of 
RKKY type\cite{eickhoff-2018,zitko2006multiple}. 

\noindent
We follow the results of the individual TBTAP molecule and fix the coupling to $\Gamma_{1}=\Gamma_{2}=20$\!~meV, which sets the two distant molecules each to a singlet ground state. An alternative analysis, in which we assume different ground states for the molecules as a result of the presence of a scanning tip, leads to equivalent results, as presented in the Supplementary Note~7. Bottom panel of Fig.~\ref{Fig3}e shows the in-gap many-body energies of SC-TIAM with respect to the ground-state energy as functions of $\zeta$. We set the direct hopping $t$ and the capacitive coupling $W$ to zero, that is, the molecules are correlated only through the substrate. The top panel shows the corresponding spin-spin correlation function $\langle\bm{S}_1\bm{S}_2\rangle$ for the ground state.

\noindent
For almost decoupled impurities ($\zeta\rightarrow0$) a combined ground state is a singlet\cite{zonda2023generalized} and the spins are very weakly correlated. There is only a single YSR at very low energies, as individual molecules are close to a QPT. With increasing $\zeta$ (decreasing distance), the YSR state splits into two that lie very close to each other; therefore, they might not be distinguishable in the STS data. 
At $\zeta\simeq0.1$ the system undergoes a QPT and the ground state changes to a doublet, allowing for a richer in-gap spectrum. Up to three pairs of YSR states are predicted for intermediate values of $\zeta$ and two pairs for large $\zeta$ (small distance). The results suggest that the ground state for both dimers is a doublet and the inner pair of YSR states in the experiment is due to the doublet-triplet transition emerging as a result of the correlations between the molecules through the surface. The YSR energies are shifted to larger values with decreasing distance, in agreement with the experimental result. At $\zeta\rightarrow 1$ the outer pair is pushed towards the edge of the gap and disappears from the spectrum.

\noindent
This analysis shows that SC-TIAM qualitatively reproduces the experimental results. However, the energies of the YSR states from the NRG do not comply with the experimental data. Nevertheless, they can be further tuned by assuming finite $t$ and $W$. Both the inter-molecular coupling $W$ and the direct hopping $t$ are expected to be significant only at small distances. Figs.~\ref{Fig3}f,g show the effect of both for $\zeta=0.87$. A hopping with energy smaller than or comparable to $\Gamma$ can fine-tune the YSR energies where the singlet state quickly approaches and finally crosses first the triplet state and then the doublet as $t$ increases. Consequently, a relatively small $t$ can switch the dimer from a doublet to a molecular singlet ground state. Conversely, as $W$ increases, the difference between the triplet and doublet state energies widens, as illustrated in Fig.~\ref{Fig3}g. This significantly shifts the position of the inner YSR peaks to higher energies, which is in compliance with experimental results. The higher weight of the inner pair of YSR states compared to the outer pair shown in the surface DOS plotted in Fig.~\ref{Fig3}h is consistent with the experimental observation. A similar approach to obtain spectra in compliance with the configuration in Fig.~\ref{Fig3}a is discussed in the Supplementary Note~8. 

\noindent
Note that both $t$ and $W$ can independently trigger a QPT from molecular doublet to the singlet ground state. At the critical point, $\langle\bm{S}_1\bm{S}_2\rangle$ changes discontinuously from positive to negative values, signaling a transition from effective ferromagnetic to effective antiferromagnetic exchange coupling. However, we must be careful with the interpretation of $\langle\bm{S}_1\bm{S}_2\rangle$ in the doublet phase. Here, the spin of the molecules is partially screened by the conducting electrons. In the idealized case, one of the spins is fully screened and the other stays unscreened, leading to a vanishing correlator. A finite value can still emerge as a result of charge fluctuations. Due to the partial screening of the spin, the correlator is strictly smaller than $1/4$, i.e., the asymptotic value of strong ferromagnetic exchange expected in the triplet state. In this respect, the observed value $\langle\bm{S}_1\bm{S}_2\rangle\approx 0.2$ is large, and therefore, we still interpret this result as an effective ferromagnetic exchange with partially screened spins.

\noindent
The spectrum of the dimer also depends on the relative orientation of the molecules. Despite a similar distance, the YSR state splits only in some cases. We can rule out the effect of magnetic anisotropy, since the isolated TBTAP$^{\bullet-}$ molecules are in spin-1/2 ground state and are free of any heavier atoms. Similar splitting of the YSR state in molecular dimers was also observed in Co phthalocyanine dimers\cite{kezilebieke2018coupled} and was attributed to ferromagnetic coupling via a phenomenological model. This conclusion is consistent with small positive values of $\langle\bm{S}_1\bm{S}_2\rangle$ in our SC-TIAM analysis. However, in order to understand the influence of the orientation of the molecules on the splitting, a detailed ab-initio analysis would be required. Admittedly, the formula that would connect $\zeta$ to the distance between the molecules and their relative orientation would be complicated. For example, the theory of two classical magnetic impurities in a superconductor predicts a complex dependence of the effective hybridization between the impurities on the distance\cite{hoffman2015impurity}. Similar effects of distance can be expected in SC-TIAM with energy-dependent $\Gamma$.

\section*{\Large \textsf{Engineering YSR states in short molecular chains.}}
\begin{figure}[!ht]
    \begin{center} \includegraphics[width=0.8\columnwidth]{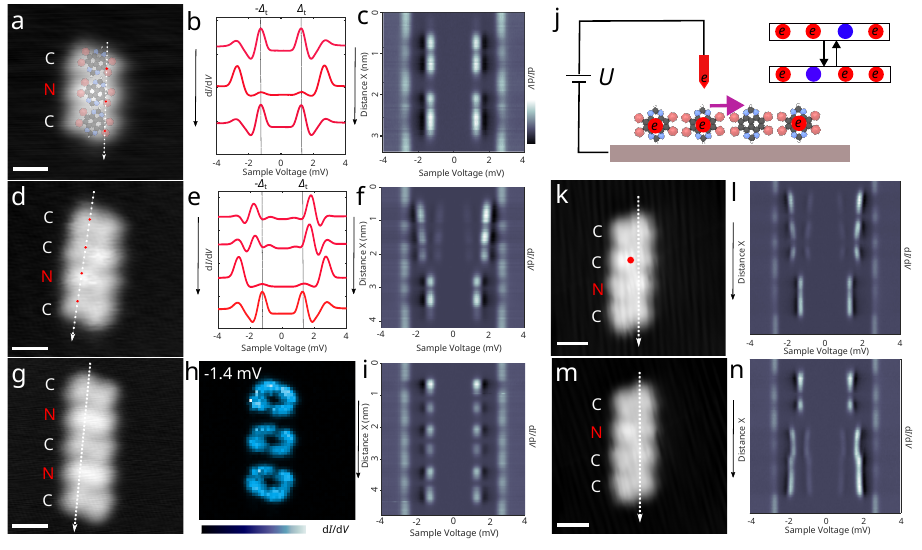} \end{center}
    \caption{\textbf{Artificial molecular chains.}
    ({\bf a}) Formation of a trimer chain via lateral manipulation, coupled by N$\cdots$H hydrogen bonding between molecules. The C (N) label denotes a charged (neutral) molecule.
    ({\bf b}) Differential conductance spectra at the positions indicated by red dots in panel {\bf a}, showing distinct electronic characteristics: a single YSR peak pair at $V_{\text{s}}=\pm 1.4$\!~mV for the end molecules and absence of YSR peaks in the central molecule.
    ({\bf c}) $dI/dV$ cross-sectional analysis along the trimer, indicated by a white arrow in panel {\bf a}. The high conductance at $V=\pm2.6$\!~mV denotes the coherence peaks of Pb(111), while the high conductance at $V=\pm1.4$\!~mV marks the YSR states.
    ({\bf d}) A tetramer chain assembled through lateral manipulations.
    ({\bf e}) $dI/dV$ spectra for the tetramer chain show a pair of YSR peaks at $V_{\text{s}}=\pm 1.9$\!~mV for the top two molecules, the absence of YSR states in the third, and a YSR peak pair at $V_{\text{s}}=\pm 1.4$\!~mV for the end molecule.
    ({\bf f}) $dI/dV$ cross-sectional analysis along the tetramer chain.
    ({\bf g}) A pentamer chain crafted by lateral manipulation.
    ({\bf h}) Grid maps of the $dI/dV$ measurements at $V_{\text{s}}=-1.4$\!~mV highlight localized electronic properties. 
    ({\bf i}) $dI/dV$ cross-sectional view across the pentamer, measured along the white arrow in panel {\bf g}.
    ($V_{\text{s}}$ = 0.1\!~V; $I_{\text{t}}=0.1$\!~nA for {\bf a}, {\bf d} and {\bf g}; spectra parameters: $V_{\text{s}}$ = 5\!~mV, $I_{\text{t}}$ = 0.1\!~nA, $A_{\text{mod}}$ = 0.03\!~mV and $f$ = 613\!~Hz). 
    {(\bf j}) Schematic representation of four TBTAP molecules on a Pb(111) surface. A negatively charged tip is positioned above one of the negatively charged TBTAP molecules. The repulsive force induces a charge transfer to the adjacent molecule, as illustrated by the arrow. The inset illustrates the equivalent states of a tetramer chain.
    ({\bf k}) and ({\bf m}) Another tetramer chain crafted to test the possibility of switching the assembly between the two possible equivalent states. When the tip is positioned at the location marked by the red circle and moved 100 pm closer to the molecule, the electron is transferred to the adjacent uncharged molecule ($V_{\text{s}}$ = 80\!~mV; $I_{\text{t}}=0.1$\!~nA). 
    ({\bf l}) $dI/dV$ cross-sectional view along the tetramer indicated by a white arrow in panel {\bf k}, showing the position of the dimer on the top two molecules.
    ({\bf n}) $dI/dV$ cross-sectional view along the tetramer indicated by a white arrow in panel {\bf m} shows the dimer localized on the bottom two molecules. Data were acquired at a sample temperature of 2.2\!~K.}
 \label{Fig4}
\end{figure}

Further control over the YSR states can be achieved in longer chains of molecules. Fig.~\ref{Fig4}a shows a molecular chain formed by three molecules with the same orientation. Tunneling spectra and a $dI/dV$ cross-section are plotted in Figs.~\ref{Fig4}b-c. They show that the first and third molecules exhibit a YSR state at $V=\pm1.4$\!~mV, similarly to the case of an isolated molecule. However, the spectrum measured over the central molecule is free of any in-gap features, except for faint peaks that can be attributed to the extension of the YSR states of adjacent molecules. This points to the loss of the radical nature of this molecule. 

\noindent
The tetramer chain was constructed by adding another molecule with the same orientation (Fig.~\ref{Fig4}d).
The tunneling spectra (Fig.~\ref{Fig4}e) and the $dI/dV$ cross-section (Fig.~\ref{Fig4}f) reveal that the top two molecules exhibit YSR states at $V= \pm1.9$\!~mV, consistent with those seen in the dimer (Figs.~\ref{Fig3}a,c). The third molecule is in a neutral state, whereas the fourth molecule behaves as isolated with YSR states near the Fermi energy. The YSR energy shows a slight difference of $V\approx0.1$\!~mV between the first and second molecules, likely due to a weak coupling between the second and fourth molecules. Note that similarly to the case of a trimer, the molecules at the end of the chain always retain their radical nature.

\noindent
Inspired by the different behavior observed in the trimer and tetramer chains, a pentamer was constructed by adding yet another molecule to the tetramer (Fig.~\ref{Fig4}g). The $dI/dV$ cross-section (Fig.~\ref{Fig4}i) shows that the only the first, third, and fifth molecule hosts a YSR state at $eV\approx\Delta$. A grid map close to the YSR position of -1.4\!~mV was measured (Fig.~\ref{Fig4}h), confirms that the spatial shape of the YSR states in the odd-numbered molecules is the same as in individual molecules (Fig.~\ref{Fig2}a).

\noindent
These results indicate that the radical character of a specific TBTAP molecule can be controlled through the interaction with other molecules. The presence of two adjacent charged molecules is required for the change of charge state, since the first and last molecules are always in the radical TBTAP$^{\bullet-}$ state. These molecules also behave as isolated with YSR states close to the Fermi energy, despite being in contact with a neutral TBTAP$^0$ molecule. This indicates that the shift of the YSR states in dimers is the result of the magnetic interaction of the spins on both molecules, as assumed by the analysis of the dimer using SC-TIAM. The loss of the charge can be explained by the presence of an additional electrostatic field induced by the two adjacent charged molecules, which shifts the SOMO energy of the molecule above the chemical potential of the substrate, leaving the molecule in the neutral state. A similar effect was also observed in TBTAP monolayers on Pb(111)\cite{liu2023gate} and in the 
TCNQ monolayers on Au(111)\cite{torrente-2012-TCNQ}. As a result of this behavior, chains of TBTAP molecules create periodic structures of charged and neutral units. Such an order can be achieved only in chains of odd lengths, as the first and last molecules are always charged. In chains of even length, this is compensated by emergence of a dimer structure on one of the ends. Such a system is in a frustrated state with two equivalent charge configurations, as the dimer can form on either end of the chain. External perturbations may induce charge transfer to an adjacent molecule in the chain and switch the system to the other energy minimum, as depicted in Fig.~\ref{Fig4}j. 

\noindent
To test this concept, we constructed another tetramer chain (Figs.~\ref{Fig4}k,m). The $dI/dV$ cross-sectional view along the chain (Fig.~\ref{Fig4}l), indicated by a white arrow in panel k, displays the dimer structure appearing on the top two molecules. When the tip is positioned at the location marked by the red circle and moved 100 pm closer to the molecule, the chain is switched to the other energy minimum and the $dI/dV$ cross-sectional view (Fig.~\ref{Fig4}n) now shows the dimer localized on the bottom two molecules, implying successful charge transfer. This switching process has been repeatedly performed (not shown), demonstrating its reversibility and the stability of the charge configuration. This suggests a possibility of encoding information in such assemblies.

\section*{\Large \textsf{Conclusions}}

\noindent
We demonstrated that metal-free TBTAP molecules on Pb(111) are easy to manipulate using the STM tip, they always adsorb to the surface in the same manner, and can be combined to create stable assemblies. Individual molecules are in a radical state with spin 1/2, which gives rise to well-developed YSR states. These states are spatially delocalized along the organic backbone of the molecule. Although other metal-free organic radicals exist\cite{mas2012attaching,zhang2013temperature,island-2017-PTM}, they do not allow for a higher level of manipulation. The individual TBTAP molecules are close to a singlet-doublet impurity QPT that can be induced by a scanning tip. The properties of this molecule can also be tuned by the presence of the second TBTAP molecule: the inter-molecular distance tunes the YSR energy, while the splitting of the YSR states into two pairs can be engineered by changing their relative orientation. Yet another type of control can be achieved in longer chains of molecules. Chains with odd and even number of constituents behave differently. Odd-numbered chains exhibit a periodic structure of YSR states localized on every second molecule, always including the ends of the chain. Chains of even length display a more complicated behavior in which a molecular dimer is formed on one of the ends. The dimer can be transferred from one end to the other by external electric field induced by the presence of the scanning tip, opening the possibility to store information in these structures. Together, the different assemblies of TBTAP molecules can be utilized as highly tunable building blocks for superconducting molecular quantum technologies.


\section*{\Large {Methods}}

{\large \textsf{Sample preparation}}\\
The Pb(111) surface (MaTeck GmbH, 99.999\%) was cleaned by cycles of Ar ion sputtering and annealing. TBTAP molecules were prepared according to the literature procedure\cite{zhou2021effect} and sublimated at $\approx 440$\!~K with the substrate kept at $\approx 100-150$\!~K. 

\noindent
{\large \textsf{Lateral manipulation}}\\
To manipulate isolated TBTAP molecules on the Pb(111) surface, we employed a lateral manipulation process commonly used to move molecules on metal surfaces\cite{liu2014,li2017,li2022surface}. Initially, we position the STM tip directly above the designated molecule using normal scanning parameters ($I=100$\!~pA, $V_\text{s}=100$\!~mV). Subsequently, we lower the STM tip to approximate position of the target molecule employing $I=3$\!~nA and $V_\text{s}=3$\!~mV, while disabling the feedback mechanism. We then moved the tip to the desired location at a speed of 200 pm/s. Finally, upon reaching the target location, we reactivated the feedback and reverted the scanning parameters to $I=100$\!~pA, $V_\text{s}=100$\!~mV. 

\noindent
{\large \textsf{STM/STS experiments}}\\
STM and STS experiments were performed with a low-temperature ($1$\!~K) Joule-Thomson STM/AFM microscope (purchased from Omicron GmbH) in ultra-high vacuum (UHV) of $\approx$ $10^{-10}$ mbar operated with Nanonis RC5e electronics. Differential conductance, $dI/dV$, spectra were recorded with a lock-in amplifier using modulation amplitudes indicated in the figure captions. The measurements were performed with a pure Pb tip to enhance the energy resolution beyond the thermal limit.

\noindent
{\large \textsf{Ab-initio calculations}}\\
We used density functional theory to find the equilibrium position of a TBTAP molecule adsorbed onto Pb(111) surface. Calculations were carried out using Turbomole 7.5.1\cite{turbomole} \texttt{ridft} code utilizing the def2-TZVP (triple-zeta) basis set and B3-LYP exchange-correlation functional with DFT-D3 (Becke-Johnson) dispersion correction. The Pb surface was modeled as a slab of three atomic layers, 60 Pb atoms in total. The system was charged by one electron to simulate charge transfer to the molecule.

\noindent
{\large \textsf{Deconvolution of the tunneling spectra }}\\
A numerical deconvolution procedure was used to obtain the surface density of states from the $dI/dV$ data. We used the maximum entropy method\cite{Jarrell-1996} implemented in a modified \texttt{ana\_cont} package\cite{Kaufmann-2023}. We used a flat default model of the same width as the input data. The optimal value of the hyperparameter $\alpha$ was obtained using the \textit{chi2kink} method\cite{Bergeron-2016}. The STM tip parameters were fitted from tunneling spectra measured on bare Pb surface separately for each data set. Additional details on the deconvolution procedure are presented in the Supplementary Note~2. The modified \texttt{ana\_cont} code is available from the authors upon request.

\noindent
{\large \textsf{Definition of the Kondo temperature}}\\
The Kondo temperature $T_\text{K}$ of the single impurity Anderson model (see Supplementary Note~3) at half-filling ($\epsilon=-U/2$) reads\cite{Hewson-1993}
\begin{equation}
k_\mathrm{B} T_\mathrm{K} = 0.29 \sqrt{\Gamma U}\exp \left[-\frac{\pi |\epsilon| (U+\epsilon)}{2 \Gamma U} \right]
=0.29 \sqrt{\Gamma U}\exp \left[-\frac{\pi U}{8\Gamma} \right].
\label{eq:TK}
\end{equation}
The value of the numerical prefactor depends on the definition of $T_\text{K}$. We use Wilson's definition for the wide-band limit as it is common in NRG, perturbation theory, and Bethe ansatz studies. A short discussion of how this definition relates to other commonly used ones can be found in Ref.~\citeonline{zitko2011kondo} and a useful table of conversions between different Kondo temperature definitions can be found in Ref.~\citeonline{turco2023accurate}. For our purposes, we only need to relate the above definition of $T_\text{K}$ to the Frota parameter $\Gamma_\text{F}$. This parameter can be extracted from the normal (non-superconducting) state zero energy Kondo anomaly by fitting the peak with the standard Frota formula\cite{frota86,frota92}
\begin{equation}
\frac{dI}{dV}\left(V\right)\propto \Ree \sqrt{\frac{i\Gamma_{\mathrm{F}}}{i\Gamma_{\mathrm{F}}+eV}}.
\label{eq:Frota}
\end{equation}
The parameter $\Gamma_\text{F}$ is tied to the half width taken at half maximum $\Gamma_\text{HWHM}$ of the peak by $\Gamma_\text{HWHM}=2.542\Gamma_\text{F}$. In experiments $\Gamma_\text{HWHM}$ is often synonymous with the Kondo temperature $T_\text{KF}$ if $T_\text{KF}\gg T_\text{exp}$ or is extracted from the Fermi liquid result,
\begin{equation}
    \Gamma_\mathrm{HWHM} = \frac{1}{2}\sqrt{(2\pi k_{\mathrm{B}}T_\mathrm{exp})^2 + 
    (2 k_{\mathrm{B}}T_\mathrm{KF})^2},
\end{equation}
if the experimental temperature $k_BT_\text{exp}$ is comparable to the width of the Kondo peak. However, Wilson's $T_\text{K}$ is related to $\Gamma_\text{HWHM}$ by $k_\mathrm{B}T_\text{K}=\Gamma_\text{HWHM}/3.7$ and consequently to the Frota fit by $k_\mathrm{B}T_{\text{K}}=0.686\,\Gamma_\text{F}$ if $T_\text{exp}$ is small. For the sake of clarity, we use the symbol $T_\text{K}^\text{FF}$ for Wilson's $T_\text{K}$ extracted by the Frota fit. This should not be mistaken for $T_\text{KF}$ as $T_\text{K}^\text{FF}=T_\text{KF}/3.7$. 

\noindent
{\large \textsf{NRG calculations}}\\
All NRG calculations were performed using the open source NRG Ljubljana package\cite{zitkoNRGljubljana}. The half-bandwidth of the conduction band was fixed to $1$\!~eV and the superconducting gap to $\Delta=1.31$\!~meV. The charging energy was set to $U=200$\!~meV, as a similar value was suggested by previous studies of TBTAP molecules on surfaces\cite{li2023strong,liu2023gate} and the local energy level to $\epsilon=-U/2$ which corresponds to a half-filled orbital. For the sake of reproducibility, we state here the NRG Ljubjana parameters used in our calculations. Detailed explanations of these parameters can be found in the package manual\cite{zitkoNRGljubljana}. Single channel calculations of the subgap states for the case of a single molecule and the dimer system with $\zeta=1$, where obtained for $\lambda=2$, $\texttt{symtype=SPSU2}$, \texttt{keepenergy=10}, \texttt{keep=6000}, \texttt{keepmin=2000}. For equivalent double-channel calculations we used $\lambda=4$, \texttt{keepenergy=6}, \texttt{keep=5000}, \texttt{keepmin=1200}. The impurity spectral functions have been obtained utilizing the \texttt{z}-averaging with $N_z=8$ and using the modified log-Gaussian kernel broadening (\texttt{smooth=newsc}) which allowed setting different values of the broadening within the superconducting gap (\texttt{omega0=1e-4} and above it (\texttt{alpha=0.15}).          

\section*{\Large Data availability}
The data supporting the findings of this study are available from Zenodo and the corresponding authors upon request.


\section*{\Large Acknowledgments}
This research was supported by the Swiss National Science Foundation (SNSF-grant 200021$\_$204053, 200021$\_$228403, 200020$\_$188445). E.M.
and S.-X.L. acknowledge the Sinergia Project funded by the SNSF (CRSII5$\_$213533). E.M. and R.P. acknowledge funding from the European Research Council (ERC) under the European Union's Horizon 2020 research and innovation programme (ULTRADISS grant agreement No 834402) and supports as a part of NCCR SPIN, a National Centre of Competence (or Excellence) in Research, funded by the Swiss National Science Foundation (grant number 51NF40-180604). C.L. acknowledges the Georg H. Endress Foundation for financial support. J.-C.L. acknowledges funding from the European Union's Horizon 2020 research and innovation programme under the Marie Sklodowska-Curie grant agreement number 847471. M.\v{Z}. and V.P. acknowledge support from Grant No. 23-05263K of the Czech Science Foundation and the EU COST action CA21144 SUPERQUMAP. Computational resources were provided by the e-INFRA CZ project (ID:90254),
supported by the Ministry of Education, Youth and Sports of the Czech Republic.

\section*{\Large Author contributions} 
E.M., R.P.\ and C.L.\ designed the experiments. P.Z., R.H., S.X.L.\ and S.D.\ synthesized the molecule. C.L. performed STM experiments. C. L. J.C.L., O.C., T.G., R.P.\ and E.M.\ analyzed the data. V.P.\ and M.\v{Z}.\ performed the theoretical calculations. C.L., V.P. and M.\v{Z}. wrote the manuscript with contributions from R.P. and E.M. All authors discussed the results and revised the manuscript.

\section*{\Large Competing interests}
The authors declare no competing interests.

\newpage

\newpage

\section*{Supplementary material}

# Supplemental Information: Nanoscale Control of Quantum States in Radical Molecules on Superconducting Pb(111)

Chao Li,[1, *] Vladislav Pokorný,[2] Martin Žonda,[3, †] Jung-Ching Liu,[1] Ping Zhou,[4] Outhmane Chahib,[1] Thilo Glatzel,[1] Robert Häner,[4] Silvio Decurtins,[4] Shi-Xia Liu,[4, ‡] Rémy Pawlak,[1, §] and Ernst Meyer[1, ¶]

[1]*Department of Physics, University of Basel, Klingelbergstrasse 82, 4056 Basel, Switzerland.*
[2]*Institute of Physics (FZU), Czech Academy of Sciences, Na Slovance 2, 182 00 Prague 8, Czech Republic.*
[3]*Department of Condensed Matter Physics, Faculty of Mathematics and*
*Physics, Charles University, Ke Karlovu 5, 121 16 Prague 2, Czech Republic.*
[4]*Department of Chemistry, Biochemistry and Pharmaceutical Sciences, W. Inäbnit Laboratory*
*for molecular quantum materials, University of Bern, Freiestrasse 3, 3012 Bern, Switzerland.*

## SUPPLEMENTARY NOTE 1: GRID SPECTROSCOPY OF INDIVIDUAL MOLECULES

Suppl. Fig. 1 provides additional details on grid spectroscopy of TBTAP molecules.

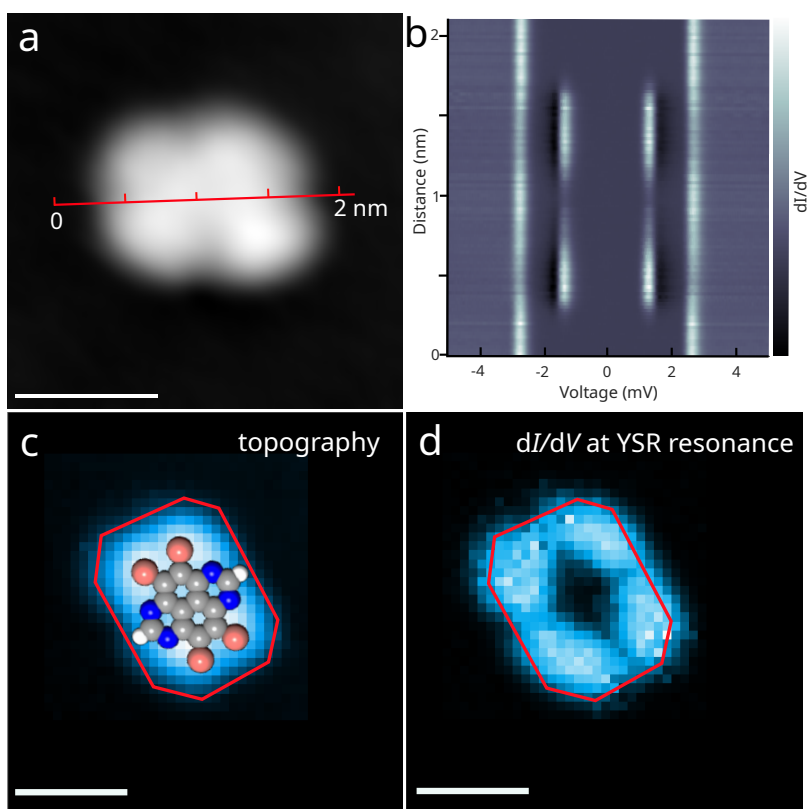

**Supplementary Figure** 1. (**a - b**) Tunneling spectra were acquired at diverse positions along the TBTAP molecule, ranging from the left side to the right side, providing a visual representation of the shape of the YSR resonances. ($V = 10$ mV, $I = 100$ pA). (**c**) Topography image obtained concurrently with the tunneling spectra. (**d**) $dI/dV$ map at the bias corresponding to the positive YSR peak, with a scale ranging from 0.0 to 1.6 nS. The region enclosed by the red line is derived from the topographic structure image area in **a**, indicating that the energy of YSR extends only over several atomic-scale distances in space. The white scale bars represents 1 nm. ($V_s = 5$ mV; $I_t = 0.1$ nA, $A_{mod} = 0.05$ mV, $f = 613$ Hz).

## SUPPLEMENTARY NOTE 2: DECONVOLUTION OF TUNNELING SPECTRA

A superconducting tip was utilized to increase the resolution of the tunneling spectra beyond the thermal limit. As a result, the measured tunneling spectra are convolutions of the sample and tip densities of states (DOS),

$$I(V) = G \int_{-\infty}^{\infty} d\omega \rho_s(\omega) \rho_t(\omega - eV)[f(\omega - eV) - f(\omega)], \tag{1}$$

where $\rho_{s/t}$ is the DOS of the sample/tip, $f(\omega)$ is the Fermi-Dirac distribution and $G$ is a constant proportional to the square of the tunnel matrix element between the initial and final state. A deconvolution routine was utilized to obtain the DOS of the sample. For the tip, we assume the BCS DOS,

$$\rho_t(\omega) = \text{sgn}(\omega) \, \text{Re} \, \frac{\omega + i\gamma_t}{\sqrt{(\omega + i\gamma_t)^2 - \Delta_t^2}}, \tag{2}$$

where $\Delta_t$ is the superconducting gap of the tip and $\gamma_t$ is a parameter which broadens the coherence peaks due to the finite lifetime of quasiparticles [1]. The differential conductance is given by

$$\frac{dI}{dV} = -eG \int_{-\infty}^{\infty} d\omega \rho_s(\omega) \left[ \frac{\partial \rho_t(E)}{\partial E} \Big|_{E=\omega - eV} [f(\omega - eV) - f(\omega)] + \rho_t(\omega - eV) \frac{\partial f(E)}{\partial E} \Big|_{E=\omega - eV} \right]. \tag{3}$$

We estimate the values of $\gamma_t$ and $\Delta_t$ from a measurement of a bare Pb(111) surface, away from the molecule. We substitute Eq. (2) for both $\rho_t$ and $\rho_s$ and assume that the values of $\gamma$ and $\Delta$ are the same for the tip and the surface. Optimal values of the parameters can be obtained using the least-squares fitting procedure. The resulting values for different experiments are included in figure captions.

With the knowledge of $\rho_t$ we rewrite Eq. (3) in a discretized form,

$$\left( \frac{dI}{dV} \right)_i = G \sum_{j=1}^{N_\omega} K_{ij} \rho_{s,j}, \tag{4}$$

where

$$K_{ij} = -e \left[ \frac{\partial \rho_t(E)}{\partial E} \Big|_{E=\omega_j - eV_i} [f(\omega_j - eV_i) - f(\omega_j)] + \rho_t(\omega_j - eV_i) \frac{\partial f(E)}{\partial V} \Big|_{E=\omega_j - eV_i} \right] \delta\omega, \tag{5}$$

$V_i$, $i = 1..N_V$ and $\omega_j$, $j = 1..N_\omega$, are discrete, equidistant voltage and energy values, $\rho_{s,j} = \rho_s(\omega_j)$ and $\delta\omega = \omega_2 - \omega_1$.

In general, such deconvolution is an ill-posed problem (the solution is not unique), since the condition number of the kernel matrix $K$ is very large and the direct inversion of $K$ is not possible (even for $N_V = N_\omega$). Usually, an approximate solution can be obtained using the Moore-Penrose pseudoinverse [2] or by assuming the functional form of the surface DOS and fitting its parameters [3]. A more reliable general solution can be obtained using the maximum entropy method (MEM). This approach was previously utilized for the analytic continuation of imaginary-time Green functions to real frequencies [4]. Instead of minimizing the misfit function,

$$\chi^2(\rho_s) = \left( \sum_i \left[ G \sum_j K_{ij} \rho_{s,j} - \left( \frac{dI}{dV} \right)_i \right] \right)^T C^{-1} \left( \sum_i \left[ G \sum_j K_{ij} \rho_{s,j} - \left( \frac{dI}{dV} \right)_i \right] \right), \tag{6}$$

where $C$ is the covariance matrix, we minimize a regularized expression $\chi^2(\rho_s)/2 - \alpha S(\rho_s)$, where $\alpha$ is an ad-hoc hyperparameter that needs to be specified outside of the method,

$$S(\rho_s) = \sum_{i=1}^{N_\omega} \left[ \rho_{s,i} - D_i - \rho_{s,i} \log \frac{\rho_{s,i}}{D_i} \right] \tag{7}$$

is the entropy and $D$ is the so-called default model which contains all the prior knowledge about the surface DOS.

There are many well-optimized implementations of the MEM algorithm. We used a modified `ana_cont` package [5] in which we implemented the kernel (5). We used a flat default model of the same width as the input data which corresponds to no prior knowledge. We assumed a diagonal covariance matrix and the standard deviation of $dI/dV$ was guessed to $10^{-13}$ S. The final results depend only very weakly on this value. The optimal value of the hyperparameter $\alpha$ was obtained using the *chi2kink* method [6]. The resulting surface DOS is nonnegative by definition and robust with regard to noise in the input $dI/dV$ data. An example of the surface DOS obtained using the MEM deconvolution is plotted in Suppl. Fig. 2. The modified `ana_cont` code is available from the authors upon request.

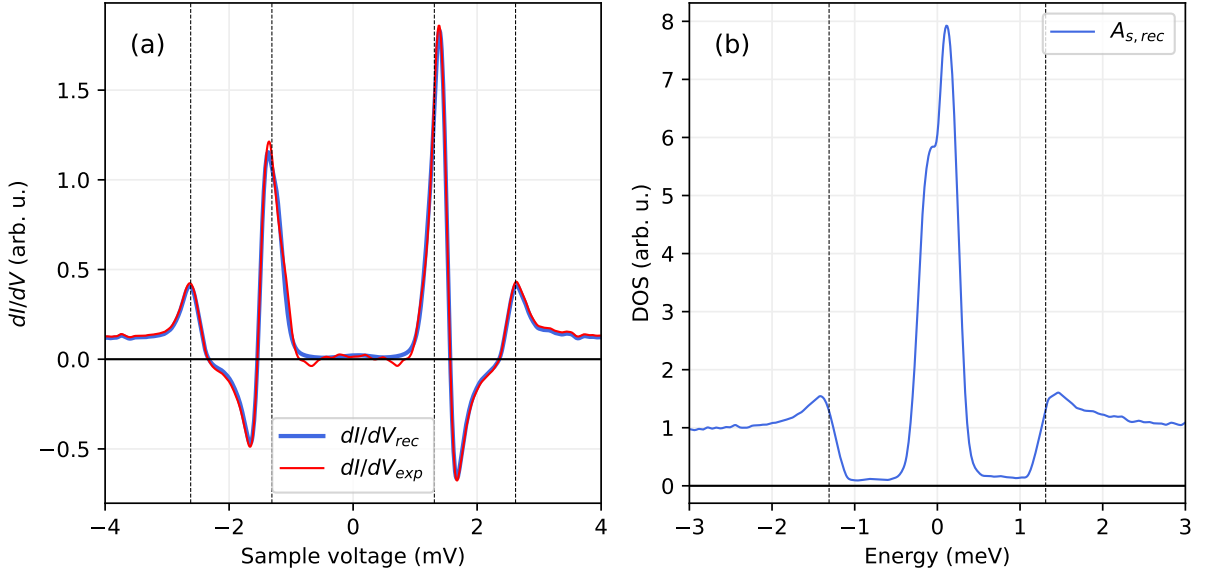

**Supplementary Figure** 2. (**a**) Differential conductance spectrum which corresponds to $h = 0$ in Fig. 2d in the main text, measured at $T_{\mathrm{exp}} = 1\,\mathrm{K}$. Red line represents the experimental result, blue line the spectrum reconstructed from the deconvolved surface DOS to check the quality of the procedure. The vertical dashed lines mark the positions of the gap edges $\pm\Delta$ and the doubled gap edges $\pm 2\Delta$. (**b**) The deconvolved surface DOS corresponding to the $dI/dV$ data in panel **a**. The parameters of the tip read $\Delta_{\mathrm{t}} = 1.31\,\mathrm{meV}$ and $\gamma_{\mathrm{t}} = 0.04\,\mathrm{meV}$.

## SUPPLEMENTARY NOTE 3: SUPERCONDUCTING SINGLE IMPURITY ANDERSON MODEL

The spectral properties of the investigated spin one-half molecule on a superconducting surface can be modeled using the superconducting single impurity Anderson model (SC-SIAM) [7]. This model describes a quantum impurity coupled to two superconducting leads, i.e. the surface (s) and the tip (t). Its general Hamiltonian can be written as

$$\mathcal{H} = \mathcal{H}_{\mathrm{1d}}^{\mathrm{imp}} + \sum_{j=\mathrm{s,t}} \left( \mathcal{H}_j^{\mathrm{lead}} + \mathcal{H}_j^{\mathrm{hyb}} \right), \tag{8}$$

where $\mathcal{H}_{\mathrm{1d}}^{\mathrm{imp}}$ describes the impurity and reads

$$\begin{aligned}
\mathcal{H}_{\mathrm{1d}}^{\mathrm{imp}} &= \epsilon \sum_\sigma d_\sigma^\dagger d_\sigma + U d_\uparrow^\dagger d_\uparrow d_\downarrow^\dagger d_\downarrow \\
&= \varepsilon \sum_\sigma \left( d_\sigma^\dagger d_\sigma - \frac{1}{2} \right) + \frac{U}{2} \left( d_\uparrow^\dagger d_\uparrow + d_\downarrow^\dagger d_\downarrow - 1 \right)^2 + \mathrm{const.},
\end{aligned} \tag{9}$$

where $d_\sigma^\dagger$ creates an electron with spin $\sigma$ on the impurity with energy $\epsilon$ and $U$ is the local Coulomb interaction (charging energy) on the impurity. For simplicity, we assume in our analysis the energy level $\epsilon = -U/2$ respectively $\varepsilon = 0$ that corresponds to a half-filled orbital.

The second term in Hamiltonian (8) describes the superconducting leads, i.e., the surface and the tip, within the BCS theory,

$$\mathcal{H}_j^{\mathrm{lead}} = \sum_{\mathbf{k}\sigma} \varepsilon_{j\mathbf{k}} c_{j\mathbf{k}\sigma}^\dagger c_{j\mathbf{k}\sigma} - \Delta_j \sum_{\mathbf{k}} \left( c_{j\mathbf{k}\uparrow}^\dagger c_{j-\mathbf{k}\downarrow}^\dagger + \mathrm{H.c.} \right), \tag{10}$$

where $c_{j\mathbf{k}\sigma}^\dagger$ creates an electron with spin $\sigma$ and energy $\varepsilon_{j\mathbf{k}}$ in the lead $j \in s, t$ and $\Delta_j$ is the superconducting order parameter. In what follows, we assume $\varepsilon_{s\mathbf{k}} = \varepsilon_{t\mathbf{k}}$ and $\Delta_s = \Delta_t \equiv \Delta$ as the surface and the tip are made from the same material. For $\Delta = 0$ we obtain the standard SIAM with metallic lead.

The last term in Hamiltonian (8) describes the hybridization between the impurity and the leads,

$$\mathcal{H}_j^{\mathrm{hyb}} = \sum_{\mathbf{k}\sigma} \left( V_{j\mathbf{k}} c_{j\mathbf{k}\sigma}^\dagger d_\sigma + \mathrm{H.c.} \right), \tag{11}$$

where $V_{j\mathbf{k}}$ is a tunneling matrix element between the lead $j =$ s,t and the impurity. In our analysis, we assume the tunnel coupling magnitudes $\Gamma_j(E) = \pi \sum_{\mathbf{k}} |V_{j\mathbf{k}}|^2 \delta(E - \varepsilon_{j\mathbf{k}})$ to be constant. The total coupling of the impurity to the leads is then $\Gamma = \Gamma_s + \Gamma_t$.

## SUPPLEMENTARY NOTE 4: KONDO TEMPERATURE IN NORMAL STATE

Suppl. Fig. 3a shows the evolution of the tunneling spectrum of TBTAP with increasing magnetic field taken at the position marked by the black dot in panel **b** at temperature $T_{exp} = 2.2$ K. At 1 T the superconductivity is completely suppressed in both the substrate and the tip. This tunneling spectrum was fitted using the Frota formula (Eq. 2 in the main text; panel **c**). The Frota fit gives the Frota parameter $\Gamma_F \doteq 0.952$ meV. The half-width of the peak then reads $\Gamma_{HWHM} \doteq 2.42$ meV and the estimated Kondo temperature is $T_K^{FF} = 7.5$ K. The spatial map of the Kondo peak is plotted in Suppl. Fig. 3d. Its shape is identical to the shape of the YSR state and has the same symmetry as the spin density of the molecule calculated using DFT (Figs. 2a-b in the main text).

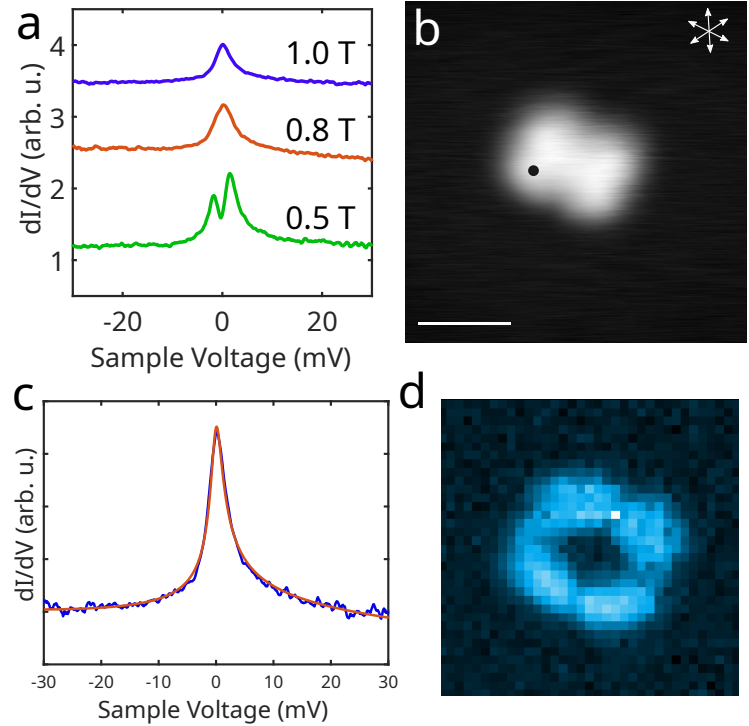

**Supplementary Figure 3.** (**a**) Differential conductance spectra obtained at the position indicated by a black dot in **b** at temperature of 2.2 K, spanning a range of magnetic fields from 0.5 T to 1 T. At a magnetic field of 1 T, the Kondo resonance appears due to the complete suppression of the superconductivity in both the Pb(111) surface and the Pb tip. The scale bar represents 1 nm. (**c**) The differential conductance spectra of Kondo resonance at 1 T (blue) and the Frota fit (Eq. 2 in the main text; red). The Frota fit yields an estimated Kondo temperature $T_K^{FF} = 7.5$ K. (**d**) A $dI/dV$ map at the bias corresponding to the Kondo peak (0 V) is the same as the distribution of YSR states in the molecule (Fig. 2a in the main text). The white scale bar in panel **b** represents 1 nm. ($V_s = 30$ mV; $I_t = 0.1$ nA, $A_{mod} = 0.03$ mV, $f = 613$ Hz).

## SUPPLEMENTARY NOTE 5: RELATION BETWEEN KONDO TEMPERATURE AND SCANNING TIP HEIGHT

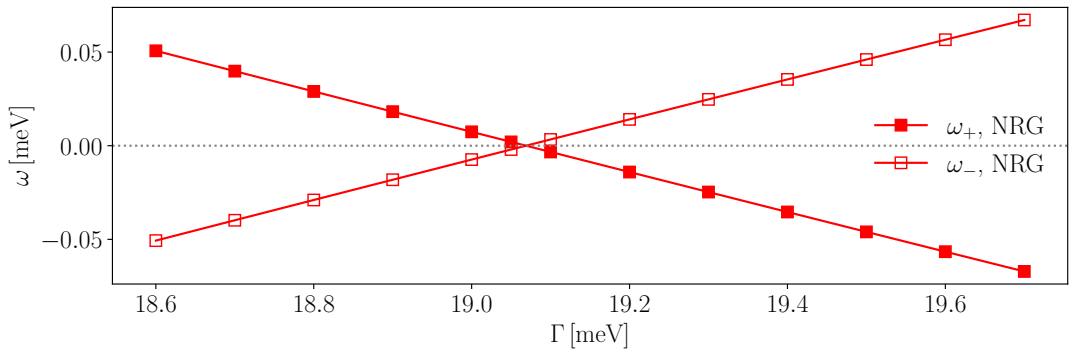

**Supplementary Figure** 4. Evolution of YSR state energy of SC-SIAM with respect to total coupling $\Gamma = \Gamma_\mathrm{s} + \Gamma_\mathrm{t}$ calculated using NRG. The model parameters are $U = 200\,\mathrm{meV}$, $\Delta = 1.31\,\mathrm{meV}$, $W = 1\,\mathrm{eV}$ and $\varepsilon = 0$.

To perform a fit of the experimental YSR energy by NRG, we need the mapping between the Kondo temperature $T_\mathrm{K}$ (or equivalently, the total coupling $\Gamma$) and the height ($z$-position) of the scanning tip $h$.

Suppl. Fig. 4 shows the NRG result on the evolution of YSR states in the vicinity of the QPT for a window of YSR energies $\omega$ reflecting the experimental values. We use model parameters $U = 200\,\mathrm{meV}$, $\Delta = 1.31\,\mathrm{meV}$, $W = 1\,\mathrm{eV}$ and $\varepsilon = 0$ and plot $\omega_\pm$ as a function of the total coupling $\Gamma$. The dependence of $\omega_\pm$ on $\Gamma$ is almost perfectly linear here and can be fitted by $\omega_\pm^\mathrm{NRG} = \pm(a + b\Gamma^\mathrm{NRG})$ with $a \doteq 2.06\,\mathrm{meV}$ and $b \doteq -108$. In principle, if all other parameters are fixed, we can extract the relation between the total coupling $\Gamma$ and the tip height $h$ as

$$\Gamma^\mathrm{exp}(h) = (\omega_a^\mathrm{exp} - a)/b \tag{12}$$

and use it for the mapping. However, there is an ambiguity here. Due to the symmetric nature of $\pm\omega_a$ it is not immediately clear whether larger $h$ means stronger or weaker $\Gamma$. As argued before [8], the STM tip exerts a force on the molecule that reflects a Lennard-Jones-like potential. The force is attractive at large tip-molecule distances and repulsive for short ones. As we lower the tip, the $\Gamma_\mathrm{t}$ increases; however, the tip also lifts the molecule from the surface which decreases $\Gamma_\mathrm{s}$. Because $\Gamma_\mathrm{t} \ll \Gamma_\mathrm{s}$ the net effect is a decrease in the total coupling $\Gamma$. It is expected that $\Gamma_\mathrm{s}$ changes exponentially with the distance [9] from the surface $z(h)$, which is a function of tip position $h$. Considering the narrow window of relevant values of $\Gamma$ shown in Suppl. Fig. 4 and neglecting the changes is $\Gamma_\mathrm{t}$ we can make a following linear approximation

$$\Gamma(h) = \Gamma_{z=0}\exp[-\alpha z(h)] \approx \Gamma_0 + \tilde{\alpha}h, \tag{13}$$

where the total coupling at $h = 0$ denoted $\Gamma_0$ and parameter $\tilde{\alpha}$ can be extracted from a fit using $\Gamma^\mathrm{exp}(h)$ from Eq. (12). This gives us $\Gamma_0 \doteq 19.7\,\mathrm{meV}$ and $\tilde{\alpha} \doteq 5.6\,\mu\mathrm{eV/pm}$. The resulting mapping to distance $h$ is then simply

$$h^\mathrm{NRG} = (\Gamma^\mathrm{NRG} - \Gamma_0)/\tilde{\alpha}. \tag{14}$$

This simple linearization is valid for $h \geq -140\,\mathrm{pm}$ as visible from Fig. 2h in the main text. For smaller distances, the potential becomes repulsive and the molecule is pushed back to the surface. Consequently, $\Gamma$ starts to grow again.

**SUPPLEMENTARY NOTE 6: SUPERCONDUCTING TWO IMPURITY ANDERSON MODEL**

We analyze the scenario of two molecules on a substrate using the superconducting two impurity Anderson model (SC-TIAM) given by Hamiltonian

$$\mathcal{H} = \mathcal{H}_{2d}^{imp} + \mathcal{H}^{lead} + \sum_{\ell=1,2} \mathcal{H}_\ell^{hyb}. \tag{15}$$

The term $\mathcal{H}_{2d}^{imp}$ describes two interacting impurities

$$\mathcal{H}_{2d}^{imp} = \sum_{\ell=1,2} \left[ \varepsilon_\ell \sum_\sigma \left( d_{\ell\sigma}^\dagger d_{\ell\sigma} - \frac{1}{2} \right) + \frac{U_\ell}{2} \left( d_{\ell\uparrow}^\dagger d_{\ell\uparrow} + d_{\ell\downarrow}^\dagger d_{\ell\downarrow} - 1 \right)^2 \right]$$
$$- t \sum_\sigma \left( d_{1\sigma}^\dagger d_{2\sigma} + \text{H.c.} \right) + W \sum_\sigma d_{1\sigma}^\dagger d_{1\sigma} \sum_\sigma d_{2\sigma}^\dagger d_{2\sigma},$$

where $d_{\ell\sigma}^\dagger$ creates an electron with spin $\sigma$ on the impurity $\ell$ with energy $\epsilon_\ell = \varepsilon_\ell - U_\ell/2$. We have found that the first line of terms in Eq. (16) is already sufficient for a qualitative explanation of the experiments. However, the interactions in the second line can be significant if the molecules are close to each other. In particular, we assume either a direct hopping (see Ref. [10]) of an electron between the impurities parameterized by $t$ or an inter-molecular density-density interaction with strength $W$.

The second term in Hamiltonian (15) describes the superconducting substrate,

$$\mathcal{H}^{lead} = \sum_{\mathbf{k}\sigma} \varepsilon_{\mathbf{k}} c_{\mathbf{k}\sigma}^\dagger c_{\mathbf{k}\sigma} - \Delta \sum_{\mathbf{k}} \left( c_{\mathbf{k}\uparrow}^\dagger c_{-\mathbf{k}\downarrow}^\dagger + \text{H.c.} \right). \tag{16}$$

The last term in Hamiltonian (15) describes the hybridization between the impurity $\ell$ and the substrate,

$$\mathcal{H}_\ell^{hyb} = \sum_{\mathbf{k}\sigma} V_{\ell\mathbf{k}} \left( c_{\mathbf{k}\sigma}^\dagger d_{\ell\sigma} + \text{H.c.} \right). \tag{17}$$

Note that in general the effect of the distance between the molecules is encoded in the complex hybridization terms $V_{\ell\mathbf{k}}$ and it also affects $t$. However, we assume the tunnel-coupling magnitudes $\Gamma_\ell(E) = \pi \sum_{\mathbf{k}} |V_{\ell\mathbf{k}}|^2 \delta(E - \varepsilon_{\mathbf{k}}) = \Gamma_\ell$ to be constant and, therefore, for the constant DOS in the substrate, the effect of the distance on hybridization is parameterized by the cross terms of the form $\Gamma_{12}(E) = \pi \sum_{\mathbf{k}} V_{1\mathbf{k}}^* V_{2\mathbf{k}} \delta(E - \varepsilon_{\mathbf{k}}) = \zeta \sqrt{\Gamma_1 \Gamma_2}$, where $0 \leq \zeta \leq 1$. Equivalently, due to the additive nature of the relevant tunneling self-energies, we can model such a system by assuming two leads as illustrated in Suppl. Fig. 5. The first impurity is coupled to the first lead by $(1-\delta)\Gamma_1$ and to the second lead by $\delta\Gamma_1$ and vice versa. The governing parameter $\delta \in \langle 0, 0.5 \rangle$ is related to $\zeta$ by $\zeta = 2\sqrt{(1-\delta)\delta}$. Here, $\zeta = 0$ ($\delta = 0$) models a situation in which molecules are not correlated through the substrate and, therefore, are effectively coupled to their own distinct reservoirs. The other limit $\zeta = 1$ ($\delta = 0.5$) described a local two-level impurity coupled to the same superconducting substrate. This means that the parameter $\zeta$ can be used to tune the inter-molecular correlations that can lead to both effective coupling as well as the effective exchange interaction of Ruderman-Kittel-Kasuya-Yosida (RKKY) type. The related spin-spin correlation function $\langle \boldsymbol{S_1 S_2} \rangle$ of the ground state can be calculated using $\boldsymbol{S}_\ell = \frac{\hbar}{2} d_\ell^\dagger \boldsymbol{\sigma} d_\ell$ where $d_\ell^\dagger = (d_{\ell\uparrow}^\dagger, d_{\ell\downarrow}^\dagger)$ and $\boldsymbol{\sigma}$ is a vector of Pauli matrices. An additional effect of the superconducting scanning tip can be approximated by varying the relevant $\Gamma_\ell$.

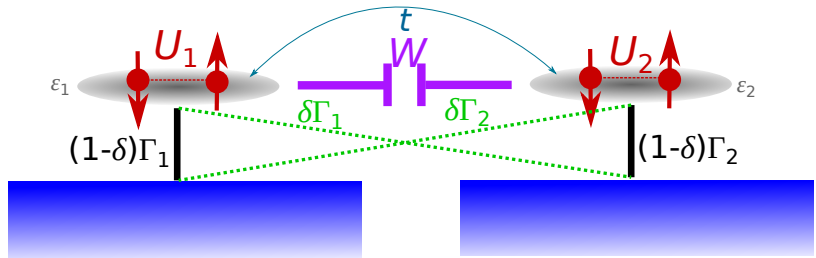

**Supplementary Figure** 5. Illustration of the SC-TIAM where the distance of the molecules is parameterized by $\delta \in \langle 0, 0.5 \rangle$. Here $\delta = 0$ (and $t = 0$, $W = 0$) describes a situation in which molecules are decoupled (large distance). In the opposite limit $\delta = 0.5$ the model is equivalent to a single, two-level impurity coupled to a single lead as a result of the additive nature of the tunneling self-energies. If the molecules are close to each other some additional interactions can play significant role, e.g., direct hopping $t$ and capacitive coupling $W$.

## SUPPLEMENTARY NOTE 7: YSR SPECTRA OF MOLECULAR DIMERS WITH DIFFERENT GROUND STATES

Suppl. Fig. 6 presents additional results on the positions of the YSR states of a molecular dimer described by SC-TIAM to complement the results in the main text. Here, one molecule is tuned to a singlet ($\Gamma_1 = 20\,\mathrm{meV}$) and the other to a doublet ($\Gamma_2 = 18.8\,\mathrm{meV}$) ground state, respectively. This change in the coupling strength can be induced by the presence of a scanning tip above one of the molecules. This contrasts with the situation in Fig. 3e in the main text, where both molecules are tuned to the same singlet ground state. The combined ground state for $t = 0$ is now always a doublet. The spectra show up to three pairs of YSR states that correspond to either doublet-triplet or doublet-singlet transitions. In this way, the situation for $\zeta > 0.1$ is equivalent to the case discussed in the main text, including the effect of nonzero direct hopping $t$ and capacitive coupling $W$.

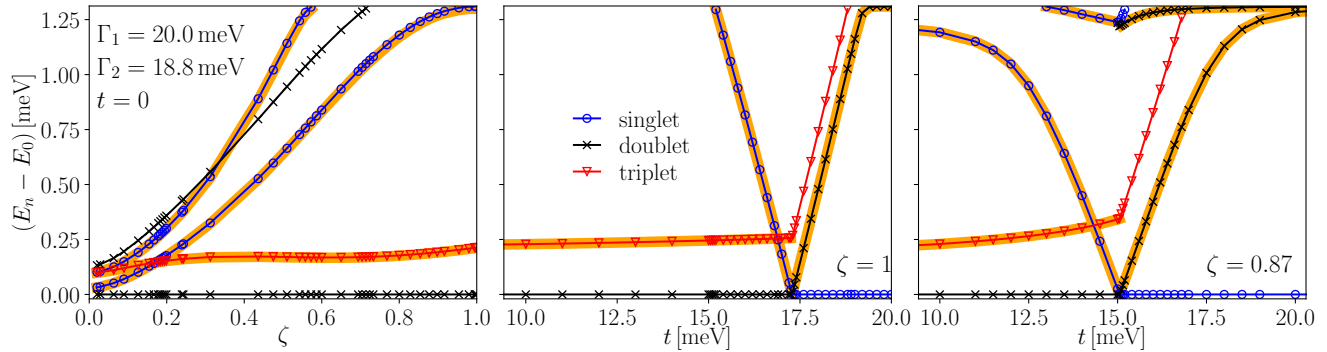

**Supplementary Figure** 6. YSR states of a dimer described by SC-TIAM calculated using NRG in the case of one molecule being in singlet and the other in the doublet ground state. The blue, black, and red symbols denote the singlet, doublet, and triplet states, respectively. Orange lines denote the actual YSR states, i.e., transitions which are not forbidden by the sum rule $\Delta s_z = \pm 1/2$.

8

# SUPPLEMENTARY NOTE 8: ADDITIONAL RESULTS ON MOLECULAR DIMERS

Here, we present additional data to support our claims on molecular dimers formed by a pair of TBTAP$^{\bullet-}$ molecules. Suppl. Fig. 7 shows the surface DOS obtained by the MEM deconvolution for the tunneling spectra presented in Fig. 3c in the main text, together with the DOS calculated using NRG for non-zero inter-impurity (capacitive) coupling $W$. Such a coupling might be sizeable in the dimer, as suggested by the results on the longer molecular chains. Tuning this coupling changes the number of YSR states, as well as their position, and allows us to simulate the DOS obtained from the experimental data. This suggests that SC-TIAM as introduced in Supplementary Note 6 is a reliable description of molecular dimers. Note that the coherence peaks at $\pm\Delta$ are not visible in the calculated DOS as a result of their small weight compared to the YSR states.

Suppl. Fig. 8 shows the evolution of the tunneling spectra and the surface DOS for a dimer with decreasing distance between the molecules to support the discussion of the effect of distance in the main text.

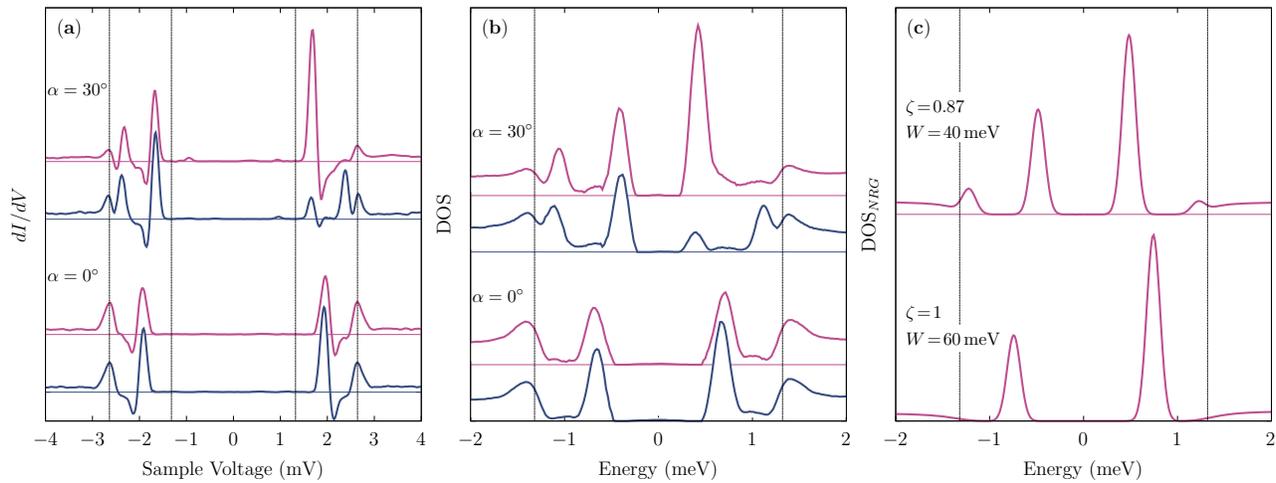

**Supplementary Figure 7.** (**a**) Tunneling spectra measured using superconducting Pb tip for the dimers formed by lateral manipulation (Fig. 3a-c in main text). (**b**) The surface density of states obtained using MEM deconvolution from tunneling spectra in panel **a**. The scanning tip parameters for the deconvolution read $\Delta_{\mathrm{t}} = 1.32\,\mathrm{meV}$ and $\gamma_{\mathrm{t}} = 0.02\,\mathrm{meV}$. (**c**) Surface DOS calculated using NRG for $U = 200\,\mathrm{meV}$, $\Gamma_1 = \Gamma_2 = 20\,\mathrm{meV}$, $t = 0$, $\zeta = 0.87$ and $W = 40\,\mathrm{meV}$ (top, $\alpha = 30°$) and $\zeta = 1$ and $W = 60\,\mathrm{meV}$ (bottom, $\alpha = 0°$), respectively. The vertical dashed lines in each panel mark the positions of $\pm\Delta$ and $\pm2\Delta$.

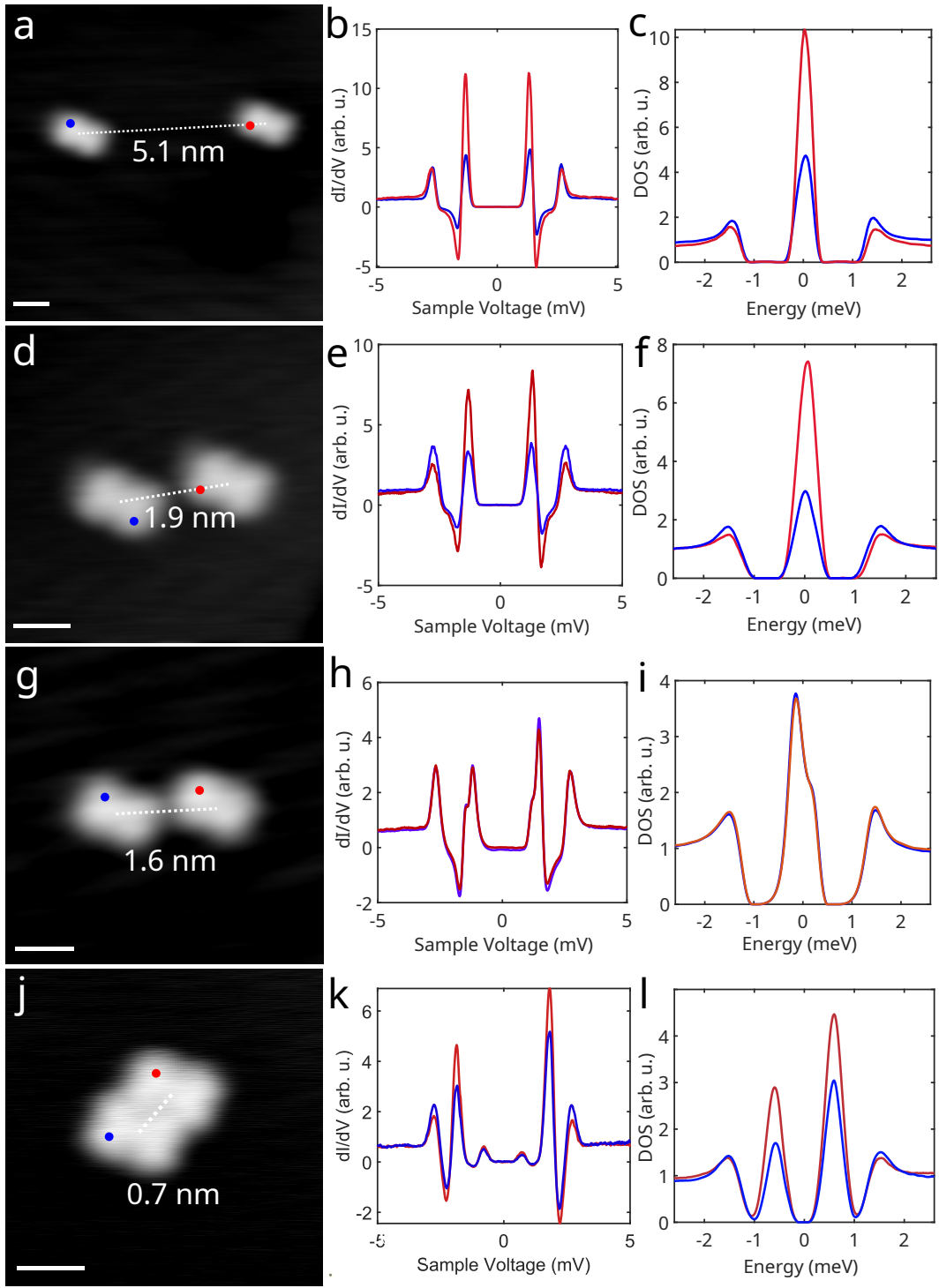

**Supplementary Figure** 8. (**a**, **d**, **g** and **j**) STM images of adjacent molecules in a relative distance of 5.1, 1.9, 1.6 and 0.7 nm, respectively. The white scale bar represents 1 nm. (**b**, **e**, **h** and **k**) Differential conductance spectra measured using a superconducting Pb tip. The spectra were obtained at the positions of the blue and red dots, represented by their respective colors, at temperature $T_{\mathrm{exp}} = 2.2$ K. The two smaller, low-voltage peaks in panel **k** are thermal images of the YSR states induced by the higher temperature which populates the states above the Fermi energy. The different intensities of the red/blue curves are a result of the fact that they were measured at different positions on the two molecules. (**c**, **f**, **i** and **l**) Surface DOS obtained from the $dI/dV$ data using the MEM deconvolution procedure. The STM tip parameters were fitted from a tunneling spectrum measured on the bare Pb surface and read $\Delta_{\mathrm{t}} = 1.32$ meV and $\gamma_{\mathrm{t}} = 0.02$ meV.

\* chao.li@unibas.ch
† martin.zonda@matfyz.mff.cuni.cz
‡ shi-xia.liu@unibe.ch
§ remy.pawlak@unibas.ch
¶ ernst.meyer@unibas.ch



\end{document}